\documentclass[printer]{aa} 
\usepackage{natbib}
\usepackage{color}
\usepackage{graphicx}
\usepackage{longtable}
\usepackage{amsmath,amsfonts,amssymb}
\usepackage[colorlinks={true},linkcolor={black},citecolor={black},
filecolor={black},urlcolor={black}]{hyperref}

\newcommand{\etal}{\text{et al. }}

\def\ion#1#2{{\rm #1}~{\rm #2}}

\begin{document}
\titlerunning{Diversity of SNe Ia determined using equivalent widths of Si
II 4000}
\authorrunning{Arsenijevic \etal 2008}

\institute{CENTRA - Centro
Multidisciplinar de Astrof\'{\i}sica, IST, Avenida Rovisco Pais, 1049
Lisbon, Portugal\\
\email{arsenije@ist.utl.pt}}

\date{Received 25 July 2008/ Accepted 17 September 2008}

\begin{abstract}
{}
{Spectroscopic and photometric properties of low and high-z supernovae Ia (SNe Ia) have
been analyzed in order to achieve a better  understanding of their diversity and to identify possible
SN Ia sub-types.}
{We use wavelet transformed spectra
in which one can easily measure spectral features. We investigate the
\ion{Si}{II} 4000 equivalent width ($EW_w\lbrace\ion{Si}{II}\rbrace$). The
ability and, especially,
the ease in extending the method to SNe at high-$z$ is demonstrated.} 
{We applied the method to 110 SNe~Ia and found  correlations
between $EW_w\lbrace\ion{Si}{II}\rbrace$  and parameters
related to the light-curve shape for 88 supernovae with available
photometry. 
No evidence for evolution of $EW_w\lbrace\ion{Si}{II}\rbrace$ with
redshift is seen. Three sub-classes of SNe~Ia were confirmed using an
independent cluster analysis with only light-curve
shape, colour, and $EW_w\lbrace\ion{Si}{II}\rbrace$.}
{SNe from high-$z$ samples seem to follow a similar grouping to nearby
objects. The $EW_w\lbrace\ion{Si}{II}\rbrace$ value measured on
a single spectrum may point towards  SN~Ia sub-classification, 
avoiding the need for expansion velocity gradient calculations.}
\end{abstract}



\title{Diversity of supernovae~Ia determined using equivalent widths of Si
II 4000}

\author{V.~Arsenijevic, S.~Fabbro, A.~M.~Mour\~ao, A.~J.~Rica da Silva}

\maketitle

\section{Introduction}

The peak luminosities of type Ia supernovae (SNe Ia) are
one of the best distance indicators at high redshifts. 
Consequently, due to the relatively small dispersion in their
light-curves, they are used for cosmological
parameter estimation (e.g.,
~\citealt{riess98,perlmutter99}).
However, some SNe seem to depart from standard behaviour and deserve 
further attention.
The most representative examples are the  
sub-luminous 1991bg-like objects, the
unusual SNe 2002ic and 2005gj\footnote{Some
authors argue that SN 2002ic may be a type Ic supernova instead of Ia (e.g.
\citealt{benetti06,wang04}), and the SN 2005gj classification as type
Ia/IIn is
still doubtful (\citealt{prieto07,trundle08}).},  and the even more
outstanding SN 2003fg (\citealt{howell06}).
The existence of a single family of SNe~Ia is thus still
debatable
(see \citealt{filippenko97} for a review on SN (in)homogeneity). Hence
 separating sub-types of SNe~Ia or attempting a
continuous parametrization of spectra and/or light-curves including the
full diversity of SNe~Ia observed up to now 
could reduce the scatter in the Hubble diagram and improve their use in cosmology.

In order to quantify the spectral differences between SNe Ia,~
 \citet{nugent95} proposed a ratio between the depths of
absorption features of \ion{Si}{II} at 5972$\textrm{\AA}$ and
6355$\textrm{\AA}$. This ratio was also found to correlate with the
absolute magnitude of SNe~Ia
and the light-curve shape parameter. We further investigate this idea
exploring a consistent method of measuring SN features on 
transformed spectra using wavelets as will be explained below.
One of the major concerns regarding the use of
SNe~Ia in cosmology is a possible systematic difference between the low-$z$
and high-$z$ samples (\citealt{blondin06,garavini07a,bronder08}).
Reasons for such evolution are generally thought to
be related to changes in metallicity or composition of the progenitor or
circumstellar medium, progenitor mass or delay times.
It is already known that progenitor age is
a relevant issue for variability in SN peak luminosity
(\citealt{prieto08,gallagher05}) and that metallicity might evolve with
redshift (see recent work of ~\citealt{ellis08} and reference therein).
We shall demonstrate that the  
equivalent width of the
\ion{Si}{II} 4000
feature, $EW_w\lbrace\ion{Si}{II}\rbrace$ defined below, coupled with light-curve
parameters, is an indicator of possible sub-classes
within SNe~Ia, consistent for both low and high-$z$ SNe.

\section{The Si II 4000 spectral feature}

The idea behind the use of spectral features to study supernovae
relies on the possibility of defining
normalized spectral ratios of these features that are common for all SNe.
It has been demonstrated
(\citealt{nugent95,folatelli04,benetti05,bronder08}) that the ratios of
the line depths or strengths are correlated with the light-curve shape.
Here we analyze the possibility of extracting from spectral
features additional information on intrinsic SN properties and their
potential to
distinguish SN sub-classes at both low and high-$z$.

For instance, the \ion{Si}{II}
$\lambda$6355 absorption feature near 6150$\textrm{\AA}$ is widely
used, being the most characteristic feature of a SN~Ia. Available
high-$z$ spectra however rarely extend to a rest-frame wavelength
of 6150$\textrm{\AA}$, thus we are forced to focus our attention on the
bluer part of the spectra, preferably the features
\ion{Ca}{II} H\&K or \ion{Si}{II} 4000 in order to compare the nearby
with the high-$z$ sample (\citealt{folatelli04}).\\
The rest-frame peak flux at the time of maximum light occurs
at about 4000$\textrm{\AA}$. At earlier epochs it is shifted slightly to
lower wavelengths, while at
later epochs the spectrum peaks at longer wavelengths. We are
particularly interested in this spectral range since almost all
high-$z$ spectra exhibit the \ion{Si}{II} $\lambda$4130 feature
blueshifted to
4000$\textrm{\AA}$, which seems to be a good choice for comparison
with low-$z$ SNe. It might serve well in distinguishing SNe~Ia
from SNe~Ib and Ic when the usual \ion{Si}{II} 6150 feature is not
available, as it is often the case with high-$z$ SNe.
The \ion{Si}{II} 4000 feature is mostly located
between 3900$\textrm{\AA}$ and 4200$\textrm{\AA}$ in the rest-frame.
Reddening effects should also be diminished since this feature is
narrow. Considering all these interesting properties, the feature is a good
candidate to be used for cosmology (\citealt{bronder08}).

The intensity of the \ion{Si}{II} 4000 feature is however often low and
contamination can be important. To study this feature we apply a simple
wavelet denoising procedure as described below. 

\subsection{Discrete wavelet transform}
\label{subsecdwt}

Wavelets are considered as a
suitable tool for studying  local properties of a signal due to their specific
structure. Indeed, their local frequency representation allows us to
process the data at different resolutions or scales. Therefore any
local defect of a signal does not affect the whole
wavelet decomposition. A local inhomogeneity
in the SN spectrum, such as remaining residuals after subtraction of
the host galaxy spectrum, sky emission, atmospheric absorption that is not 
completely removed in the reduction process, or any calibration or
instrumental undesirable effect, affects just the coefficients in
the wavelet decomposition in the small region where it appears.\\

To provide a homogeneous sample of data,
we use spectra
in the interval $3400\textrm{\AA}-7000\textrm{\AA}$. Other choices of
intervals do not make much difference due to the local character
of the wavelet transform. All spectra
were deredshifted first. Mallat's pyramid algorithm
(\citealt{mallat89}) is then
performed to obtain the Discrete Time Wavelet Transform ($DTWT$) using
Daubechies' extremal phase wavelets with 4 vanishing moments (D8) as a
basis.

\begin{figure}[!htb]
\begin{center}
\begin{minipage}[c]{0.8\linewidth}\includegraphics[width=\linewidth,
angle=-90]{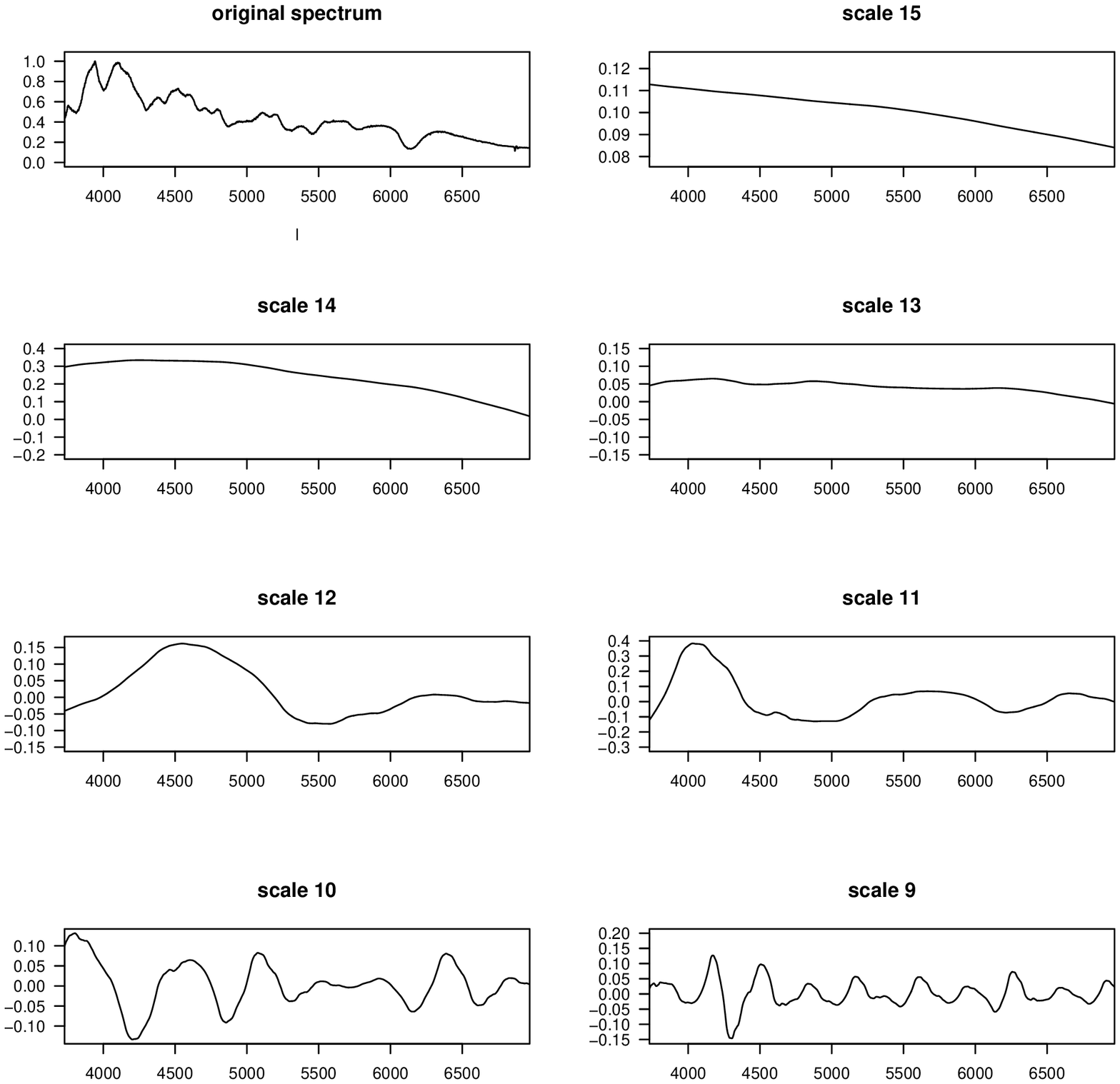}
\end{minipage}
\begin{minipage}[c]{0.8\linewidth}\includegraphics[width=\linewidth,
angle=-90]{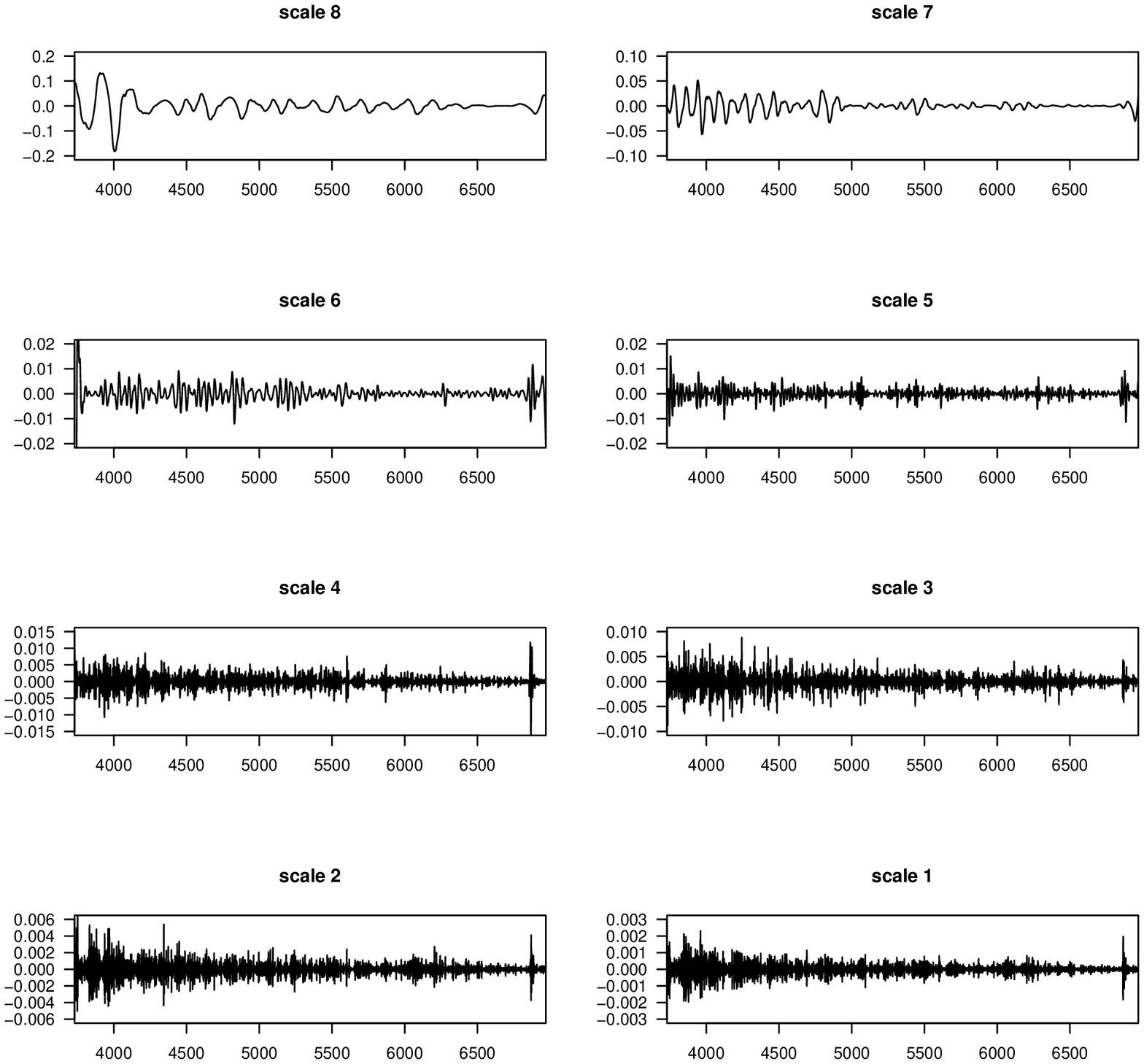}
\end{minipage}
\caption[]{\label{spectrum94dm02}{\small\sl 
Inverse wavelet reconstruction of separate scales of the wavelet
decomposition of a spectrum of SN 1994D. 
The sum of contributions from scales 7-14 is used to construct the wavelet 
transformed spectra, $F_w(\lambda)$, used in this work.}}
\end{center}
\end{figure}

The number of decomposition levels refers to the number of levels of
smoothed data and wavelet coefficients. The coefficients at the lower
levels carry ``high-frequency'' information. 
If a spectrum consists of $N=2^m$ discrete points (which can be assumed
without loss of generality),  
there will be 
$m$ wavelet coefficient bands 
and a scaling value, whose inverses are given in the
decomposition shown in Fig.~\ref{spectrum94dm02} (see ~\citealt{nason94}). 

It is often reasonable to assume that
only a few large coefficients contain relevant information about
the underlying signal, while small wavelet coefficients, especially 
from low scales, can be attributed to noise or any other undesirable high
frequency feature. Thus this part of spectra, the scales 1-6 (see
Fig.~\ref{spectrum94dm02}), will be excluded from our considerations.
This step assures that after taking the inverse $DTWT$ of such
a spectrum, the resulting wavelet transformed
spectrum becomes much 
smoother, and one can easily measure the features.

The analysis of the wavelet power spectrum\footnote{Obtained by summing
the squares of the wavelet coefficients, divided by the number of
corresponding coefficients for each scale.}
revealed that scale 15 shows much greater discrepancy among SNe, even
for spectroscopically similar events. For this reason, scale 15 will
not be included in the wavelet reconstruction of the spectra. We
can support this step by the following
technical explanation: the contribution from scale 15, shown in
Fig.~\ref{spectrum94dm02}, is obtained by the inverse wavelet transform
of a scaling constant, equal to the sample mean multiplied by the square
root of the number of original data points, and the featureless inverse of
the wavelet coefficient from the largest scale. The subtraction of the
largest
scale, as also mentioned in \citet{starck97}, does not significantly
deform the features of the spectrum.

After removal of scales 1-6 and scale 15 in the wavelet space, we apply
the inverse $DTWT$. We treat all SNe~Ia the same way to obtain a
homogeneous sample of normalized wavelet transformed
spectra. 
One could also apply a standard (soft) thresholding, but an optimal choice
for a threshold value that cuts off all smaller wavelet coefficients and is
consistent for whole variety of SN spectra is a complex issue and we have not 
attempted its implementation. Most spectral features analyses include
a boxcar or Gaussian
smoothing. Residual sky lines or galactic emission lines are thus also
smoothed leading to possible misestimations of feature bounds and extrema.
Our approach greatly reduces this effect. 

\subsection{Determination  of
\texorpdfstring{$EW_w\lbrace\ion{Si}{II}\rbrace$}{}}

In order to quantify SN~Ia diversity with spectral features, rather
than looking at deviation from the average SN~Ia spectrum
(\citealt{james06}), some authors use a pseudo-equivalent-width ($EW$)
(\citealt{hachinger06,garavini07a,bronder08}). Here, we apply the same
definition but on the wavelet transformed spectra:

\begin{equation}
 EW_w = \sum_{i=1}^{N}\frac{F_w^c(\lambda_i)-F_w(\lambda_i) } {
F_w^c(\lambda_i)
}\Delta\lambda ,
\label{ewdef}
\end{equation}
where $F_w(\lambda)$ is the transformed SN flux
after the wavelet procedure that has been 
performed on SN spectra, as described in \ref{subsecdwt} (therefore the subscript $w$); 
$F_w^c(\lambda)$ stands for the pseudo-continuum of the transformed flux,
$N$ is the number of data points;
the bounds of the feature are $\lambda_{\rm min}\!=\!\lambda_1$ and 
$\lambda_{\rm max}\!=\!\lambda_N$. 
\\

$EW_w\lbrace\ion{Si}{II}\rbrace$ is defined as the $EW$ of
 \ion{Si}{II} $\lambda 4130$ feature on 
transformed spectra,
\begin{equation*}
EW_w\lbrace\ion{Si}{II}\rbrace \equiv 
EW_w\lbrace\ion{Si}{II}\; \lambda
4130\rbrace .
\end{equation*}

To determine the pseudo-continuum bounds, we use a multiple Gaussian
peak-fitting semi-automated routine. All the pseudo-$EW$ definitions used
in the literature, including ours, depend on the accuracy of the
pseudo-continuum bounds, and therefore on the spectrum
signal-to-noise ratio (S/N).
The traditional definition of $EW$ does alleviate the dependency on S/N, 
but requires the full spectrum continuum to be well defined, which is
more difficult for the SNe~Ia. We are currently investigating this
subject. We elaborate on the S/N dependency
in Sect.~\ref{systerrors}. We also note that our wavelet-based method 
for the removal of high-frequency features adds extra robustness to the
pseudo-$EW$ estimation.

\begin{figure}[!htb]
\centerline{
\includegraphics[width=7cm,angle=-90]{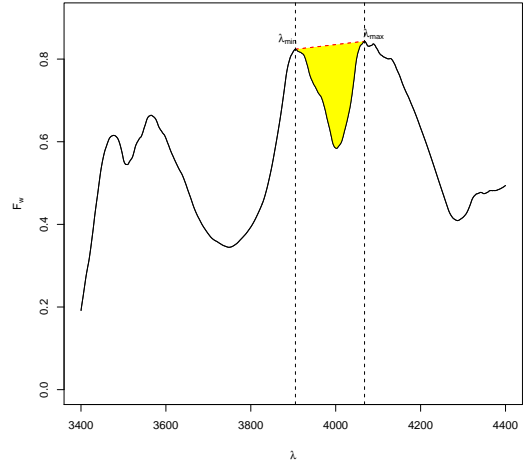} }
\caption[]{\label{rcsidef}{\small\sl Illustration of
the $EW_w\lbrace\ion{Si}{II}\rbrace$ 
definition 
on a wavelet transformed spectrum, $F_w(\lambda)$,
of SN 1994D.}
\bigskip}
\end{figure}

The $EW_w\lbrace\ion{Si}{II}\rbrace$ evolves smoothly with the SN epoch, as
we verified on
a spectral series template given by the SALT2 model (\citealt{guy07}).
We use the same prescription as ~\citet{bongard06} to obtain an
estimate of $EW_w\lbrace\ion{Si}{II}\rbrace$ at $t=0$ at maximum in $B$:
for a single epoch spectrum, we use the actual value (but increase the
error), for two epoch spectra a straight line fit, and for  multi-epoch
spectra a quadratic polynomial fit.

We also estimate the statistical
error made by the determination of extrema of the
specific features considered, as both the
wavelet transformed flux and
the pseudo-continuum come with uncertainties that affect the
calculated value of $EW_w$. 

Standard error propagation, assuming small enough errors, applied on
Eqn.~\ref{ewdef} leads to:
\begin{equation*}
 \sigma_{\rm{stat}}=\left[
\sum_{i=1}^{N}\left(\frac{\sigma_{F_w}^2(\lambda_i)}{{F_w^c}^2(\lambda_i)}
+\frac {F_w^2(\lambda_i) } {{F_w^c}^4(\lambda_i)}\sigma_c^2(\lambda_i)
\right)\left(\Delta\lambda\right)^2 \right]^{1/2},
\label{sigmacalc} 
\end{equation*}
where 
$\sigma_{F_w}$  consists of the inverse
$DTWT$ of the low scale coefficients from the wavelet decomposition 
that have been subtracted; 
$\sigma_c(\lambda)$ is obtained from the error of a straight line fit that
defines the pseudo-continuum.\\

When we have multi-epoch spectra, $EW_w\lbrace\ion{Si}{II}\rbrace$ is
estimated for each of the epochs, and the error-weighted polynomial fit
returns the propagated $\sigma_{\rm{stat}}$. %
In the more difficult but common cases of single epoch spectra, we add
to $\sigma_{\rm{stat}}$ an error floor computed as the maximum dispersion
observed on the epoch of the given spectrum. Typically we add
$\sigma_{\rm{floor}}=1.7$.\\

The dispersion seen in wavelengths of the feature bounds can
be partly explained by redshift uncertainty, typically of ~3-30
$\textrm{\AA}$ (respectively for $\Delta z$ of 0.001 and 0.01) at a
redshift of 0.5.
However, no offset in $\lambda_{\rm min}$ or $\lambda_{\rm
max}$ was found between good quality spectra and reduced S/N data. The
major reason for this is the use of the specific wavelet transform we
perform on SN spectra to obtain the spectral function  on which $EW_w$s are
measured. The main consequence is that
statistical uncertainties in the high-$z$ set are generally higher
due to lower S/N.

\subsection{Systematic errors}
\label{systerrors}

To test the robustness of the $EW_w\lbrace\ion{Si}{II}\rbrace$
measurements and quantify various systematic errors, we ran a set of
simulations taking very high S/N low-$z$ spectra on top of which we added
extra contributions (for instance sky and/or galaxy) and 
Gaussian noise with a variable standard deviation.

\begin{figure}[!htb]
\centerline{\
\includegraphics[width=8cm,angle=-90]{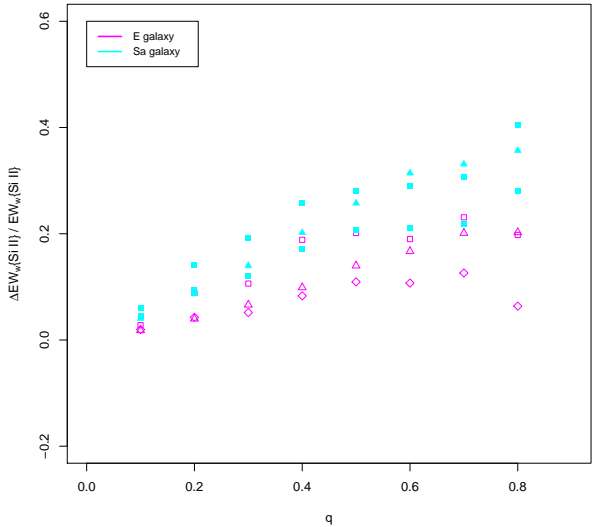} }
\caption[Relative error as a function of host
contamination]{\label{galcontamount}{Relative error of the measured 
$EW_w\lbrace\ion{Si}{II}\rbrace$ as a function of $q$, the amount of host
contamination for elliptical (open symbols) and spiral (filled symbols)
galaxies for 3 low-$z$ SNe, 1981B, 1994D and 2002bo (given as triangle,
diamond and square symbols respectively).}
\bigskip}
\end{figure}

One source of systematic errors is the host galaxy
contamination. Very often, the light from the host galaxy is not perfectly
removed and the remaining light affects the estimation of equivalent
widths. We tested different amounts of galaxy contamination, up to 80\%
of the total integrated SN flux from 3400-7000$\textrm{\AA}$. The
simulations
using template galaxy spectra of Hubble types E and
Sa indicate that increasing galaxy contamination implies lower values
of the measured $EW_w\lbrace\ion{Si}{II}\rbrace$, as found in
~\citet{garavini07a, bronder08}. 
In Fig.~\ref{galcontamount} we illustrate the correlation between
increasing galaxy contamination level and the
relative decrease in $EW_w\lbrace\ion{Si}{II}\rbrace$. 

Our results with a few SNe show a scatter
that is generally of less than 20\% for SNe in elliptical galaxies, even
for contaminations of 50-80\%, although it becomes greater for objects
hosted in spiral galaxies, but never exceeds 40\% of the true value.
It may happen, however, that extremely high host-galaxy
contamination distorts significantly the feature we measure, as in
the case of the SN 2002bo spectrum with 80\% elliptical galaxy
contamination (see~Fig.~\ref{galcontamount}).
Recall that ~\citet{bronder08} make a selection cut when
galaxy contamination is found to be more than 65\% since it can produce a
change of up to 100\% in the $EW$ value. 

We expect that at low signal-to-noise ratios, spectra do not exhibit so
clearly the 
\ion{Si}{II} 4000 feature, making it difficult to measure
accurately and possibly introducing a slight misestimation. 
Varying the noise in our simulations then
measuring $EW_w\lbrace\ion{Si}{II}\rbrace$ and
corresponding error on each spectrum  suggests that no
significant bias is present in the $EW_w\lbrace\ion{Si}{II}\rbrace$
values.%

We also checked our hypothesis of the small effect of reddening 
on the \ion{Si}{II} 4000 feature. Further simulations using the extinction
law of \citet{cardelli89} with $R_V=3.1$
have shown that the
relative error of $EW_w\lbrace\ion{Si}{II}\rbrace$ measured on spectra
with/without reddening
correction never exceeds 0.5\%, while for $EW_w$ of \ion{Ca}{II}
H\&K  reaches the order of 1\%. This fact supports
the choice of the \ion{Si}{II} 4000 feature instead of broader ones, such
as \ion{Ca}{II} H\&K or \ion{Mg}{II} 4300.
Other extinction laws based
on studies of dust properties in the Magellanic Clouds give similar
results for the wavelength range under consideration (see, e.g.
\citealt{pei92, weingartner01}) and thus were not further investigated at
this stage.

The contamination from residual sky line subtraction was also studied. To
achieve this, we added to our set of several low-$z$ SNe, which we redshifted
at $z=0.5$, a fiducial sky spectrum and Gaussian noise. We verified that
our $DTWT$ procedure properly removes these lines, leaving
$EW_w\lbrace\ion{Si}{II}\rbrace$ unmodified. 

In Table~\ref{errorsummary} we
summarize all the measurement errors mentioned above.\\
\begin{table}
\centering
\begin{tabular}{|l|l|} \hline \hline
 Source & Size \\ \hline\hline
statistical error & $15-27\%$ (low-$z$ vs high-$z$) \\
galaxy contamination & $< 20-40\%$ (E vs S type)\\
low S/N & $<5\%$ for S/N $\geq$5 \\
reddening & $<0.5\%$ \\
high frequency feature residuals & $<0.5\%$ \\
\hline
\noalign{\smallskip}
\multicolumn{2}{c}{ }\\
\end{tabular}
\caption[]{Summary of the error budget for the
$EW_w\lbrace\ion{Si}{II}\rbrace$ measurements.}\label{errorsummary}
\end{table}

\section{Data sets}

We then  applied our measurement procedure to a set of published SN~Ia
data. The main selection criterion  was the detection
of the \ion{Si}{II} spectral feature we are interested
in. It thus depends on wavelength
and epoch coverage. Tests performed on SN spectral templates indicate
that it should be present within an interval of $[-15,15]$ days relative
to 
$B$-maximum. 
All selected SN spectra  have epochs within an interval of $\pm 9$ days
around the maximum in $B$ in the rest-frame.

Furthermore, if
the presence of strong host galaxy contamination was noticed resulting
in severe misestimation of \ion{Si}{II} 4000 feature, that spectrum
was rejected.
We mainly selected SNe with published photometry in order to check for
correlations between $EW_w\lbrace\ion{Si}{II}\rbrace$ and  light-curve
parameters.

\subsection{Low-\texorpdfstring{$z$}{} sample}

The analysis is applied to a sample of 35 local SNe taken from
the literature and presented in Table~\ref{nearbylistaa2}. 
The spectroscopic data are taken mainly from public archives:
SUSPECT\footnote{\protect\url{
http://bruford.nhn.ou.edu/~suspect/index1.html}}, CfA Supernova
Archive\footnote{\protect\url{
http://cfa-www.harvard.edu/oir/Research/supernova/SNarchive.html}} and 
SUSPEND\footnote{\protect\url{
http://www.nhn.ou.edu/~jeffery/astro/sne/spectra/spectra.html}}.
Spectra of  1991T and 1991bg-like SNe, also of other peculiar SN 
events were also considered.
$EW_w\lbrace\ion{Si}{II}\rbrace$ is measured on 124 spectra of low-$z$ SNe.

\subsection{High-\texorpdfstring{$z$}{} sample}
The principal
difficulty when studying  high-$z$ SNe, 
besides much noisier spectra compared to the low-$z$ events, 
is the wavelength coverage
spanned by spectral observations resulting in the lack of major
spectral features, like \ion{Si}{II} 6150. 
However this is not the case with the \ion{Si}{II} 4000 
feature. 

From the available set of high-$z$ SuperNova Legacy Survey (SNLS) data, 
we used 26 SNe with spectral epochs within $\pm 9$ days relative
to maximum in $B$ that have been published in ~\citet{howell05}. 
This SN sample is distributed over $0.337 < z < 1.01$ (see
Table~\ref{salt2list}). 
Usually there is only one spectrum for each high-$z$ supernova.
We note that the host galaxy spectra were
not subtracted.

High-$z$ spectra from the first two years of the ESSENCE
project (see, e.g.,
\citealt{matheson05,miknaitis07}) were checked;  among these, 26 spectra
that show the presence of the \ion{Si}{II} 4000 feature
were used.
No attempts to subtract the host galaxy from the spectra were made, but
~\citet{matheson05} did not report any strong contamination within
this selected sample.

The Supernova Cosmology Project (SCP)
high-$z$ spectra used in this work were provided by \citet{hook05} and
\citet{lidman05},
as seen from Table~\ref{salt2list}. 
Measurements were performed on 6 and 10 SN spectra respectively.

We use also 7 spectra from High-$z$ Supernova Search Team (HzSST) that
were published in ~\citet{tonry03}. Spectra from ~\citet{hook05} and
~\citet{tonry03} were corrected for host galaxy light.\\

Light-curve data come from
~\citet{astier06,miknaitis07,perlmutter99,tonry03,kowalski08}.

\section{\texorpdfstring{$EW_w\lbrace\ion{Si}{II}\rbrace$}{} properties}
\label{subsec:sprat}

Armed with $EW_w\lbrace\ion{Si}{II}\rbrace$ values for 32 low-$z$ and  
75 high-$z$ SNe~Ia, we searched for correlations with light-curve
parameters.
 In order to apply a consistent procedure for all SNe, 
we fitted the available light-curves with the SALT2 model.
We ended up  with 30 low-$z$ and 58 high-$z$ SNe, 
for which $EW_w\lbrace\ion{Si}{II}\rbrace$ and SALT2 parameters were
calculated,  
 as can be seen in Tables~\ref{nearbysalt2} and ~\ref{salt2list}. We also tried our
analysis with published parameter values from the MLCS2k2
(\citealt{jha07}) and SALT (\citealt{guy05,kowalski08}) fitters.

\subsection{Relation with the light-curve parameters}

\begin{figure}[!htb]
\begin{center}
\begin{minipage}[c]{0.90\linewidth}\includegraphics[width=\linewidth,
angle=-90]{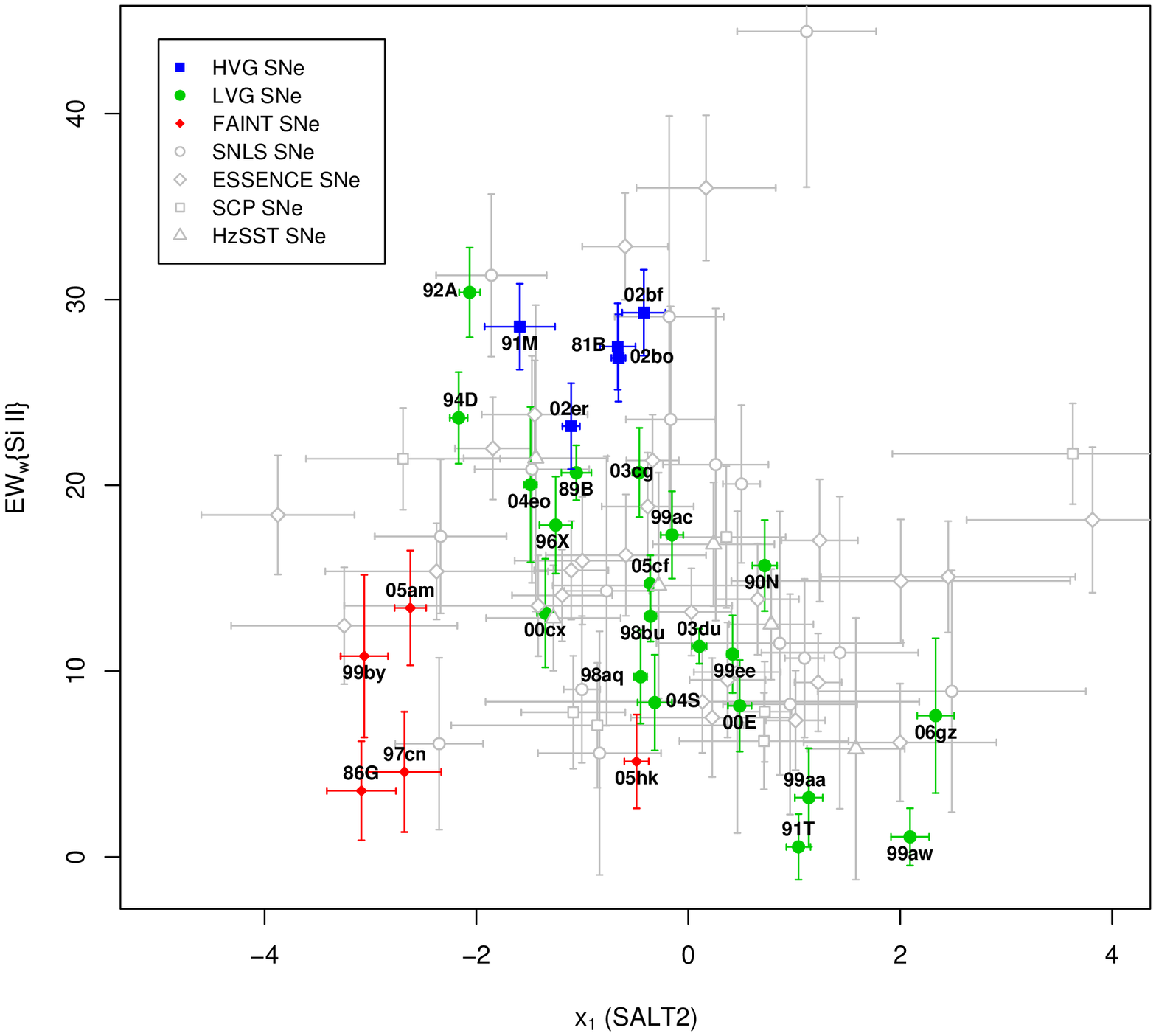}
\end{minipage}
\begin{minipage}[c]{0.90\linewidth}\includegraphics[width=\linewidth,
angle=-90]{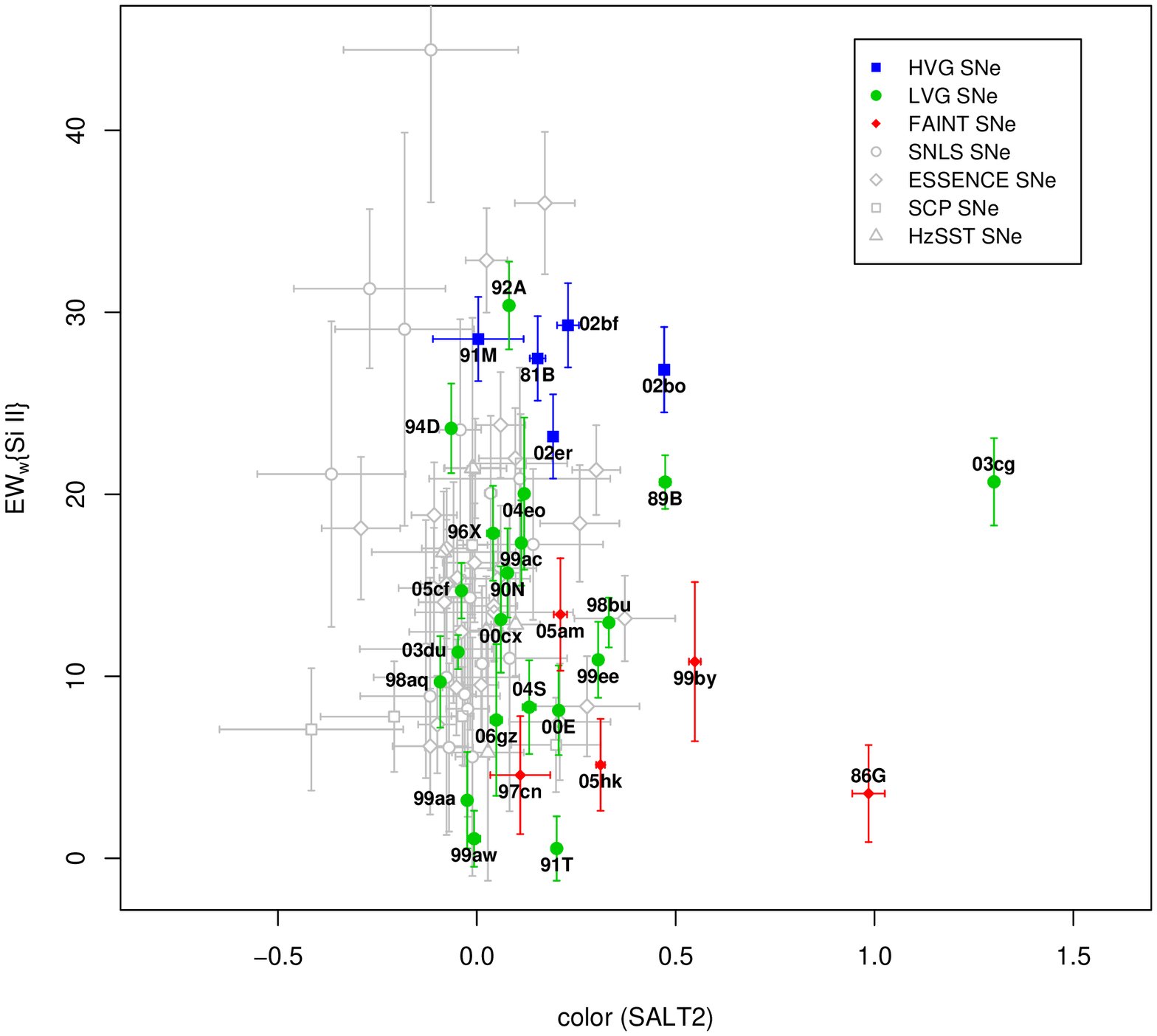}
\end{minipage}
\caption[]{\label{rcsix1}{\small\sl $EW_w\lbrace\ion{Si}{II}\rbrace$
versus $x_1$ (top) and colour (bottom) estimated using SALT2 including
high-$z$ SNe. The colours chosen for different
sub-classes are the same as in ~\citet{benetti05,pastorello07b}:
blue filled squares for HVG SNe, light green filled circles for LVG
SNe and red symbols for FAINT SNe.}
\bigskip}
\end{center}
\end{figure}

We first checked the correlation between $EW_w\lbrace\ion{Si}{II}\rbrace$ and
$x_1$ (SALT2 width parameter), illustrated in the top panel of
Fig.~\ref{rcsix1}. The $x_1$ parameter measures the
departure of the stretch of a SN from the average value of the
training sample (in standard deviations); its average value is adopted to
satisfy $<x_1>= 0$ and $<x_1^2>= 1$ (\citealt{guy07}).
~\citet{nugent95} and \citet{hachinger06}
found a strong correlation between the ratio of the depth
of the absorption features of \ion{Si}{II} at 5972 and 6355$\textrm{\AA}$,
$\cal{R}(\ion{Si}{II})$, and
$EW\lbrace\ion{Si}{II \lambda 5972}\rbrace$  and $\Delta m_{15}$. 
We confirm a similar correlation for low and high-$z$ SNe using  the \ion{Si}{II} 4000 feature and
a different light-curve model (SALT2). 

We noticed however several SNe departing from the main trend. A similar
behaviour was highlighted by \citet{benetti05} 
using velocity gradients inferred from the measured
blueshift of the \ion{Si}{II} absorption feature at 6355$\textrm{\AA}$. The
authors identified
sub-classes of SNe in the low redshift sample termed: high velocity
gradient (HVG), low
velocity gradient (LVG) and FAINT SNe. The latter group
includes those SNe that are found to have similar brightness as that of SN
1991bg.
Other authors confirmed the same sub-classifications 
(see for instance \citealt{hachinger06,pastorello07b}).
These authors also use the expansion velocity 
gradient of the \ion{Si}{II} $\lambda 6355$ feature to classify SNe,
therefore multi-epoch spectra are required. 
A similar grouping was found by \citet{branch06} studying the equivalent 
widths of \ion{Si}{II} $\lambda 6355$ and $\lambda 5972$ features. 
These approaches however are hardly applicable to high-$z$ SNe due to the wavelength coverage.  

We distinguished exactly the same three sub-classes
applying a hierarchical cluster analysis in a much smaller parameter space,
using only $EW_w\lbrace\ion{Si}{II}\rbrace$ with a SALT2  light-curve
shape, $x_1$, and colour parameter.

We emphasize as an advantage of our method the
ability  to identify the 3 sub-classes 
with only one spectrum relatively close to $B$-maximum.
If we remove the
$EW_w\lbrace\ion{Si}{II}\rbrace$ parameter from the cluster
analysis we are not able to find the three clusters of SNe. 
We also ran the cluster analysis on  low-$z$ SNe using other
estimates of light-curve width such as $\Delta m_{15}$, stretch, $\Delta
(\rm{MLCS})$ and identified the same three sub-classes.

It is known that for HVG\footnote{This group corresponds to the broad-line
group in the classification of ~\citet{branch06}.} SNe the \ion{Si}{II}
$\lambda6355$ line evolves
rapidly (\citealt{benetti05}). These SNe generally have larger
photospheric velocities than SNe with a slower evolution of the
\ion{Si}{II} $\lambda6355$ feature, such as the LVG SNe. 
This latter  group includes both
normal SNe~Ia and the brightest ones. Again, in the
classification of ~\citet{branch06}, the LVG group corresponds
to core-normals and shallows. However, SNe from both HVG
and LVG groups have similar maximum luminosities. 
The difference in
photospheric velocity comes as a consequence of the 
difference in photospheric temperature; the HVG have a lower
temperature compared to LVG SNe. 

There are also a few SNe,
namely 1983G, 1984A, 2002bf and 
2004dt, that show similar behaviour to that
of SN 2002bo, a well studied HVG supernova. 
The similarities between these SNe were also
pointed out by ~\citet{altavilla07} and \citet{leonard05}. 
All of them belong to the group of SNe with unusually high photospheric velocities.
In addition,  2004dt and 2002bf 
are both highly polarized. Spectropolarimetry can provide, in
general, a probe of supernova geometry;  greater divergence from
spherical symmetry normally causes a higher polarization,
but the latter may also be caused by clumping
(see \citealt{wangwheeler08} for the latest review).
Furthermore, ~\citet{wangbaadepatat07} found significant
peculiarity
of SN 2004dt comparing the degree of polarization across the
\ion{Si}{II} $\lambda$6355 line and light-curve decline
parameter $\Delta m_{15}$.\\

The difference between HVG and LVG SNe
can also be studied in objects that have similar decline rates, as shown
in ~\citealt{tanaka08} for SNe 2002bo and
2001el. They conclude that burning in LVG is less powerful than in
HVG SNe, thus there is a difference in kinetic energy of the ejecta
(of the order of $\sim$2\%)\footnote{Although this difference is too
small to explain the spectral diversity between the HVG and LVG SNe.}.
This difference in kinetic energy  implies that 
HVG SNe light-curves are narrower than  LVG SNe ones.
Even if we assume that it is the only difference between these two groups
of SNe, this might cause an intrinsic dispersion in the
luminosity/light-curve shape relation. 

As shown in Fig.~\ref{rcsix1}, we also find 4 
events considered as non-standard in
~\citet{pastorello07b,branch06,hachinger06}, namely SNe 1989B,
1991M, 1992A and 2004eo. 
Among them, SN 1991M is labeled as HVG SN,
while the others are shown as LVG objects. 
We may add to this list an HVG SN 2002er that has been found to have
properties common to both HVG and LVG SNe, like a lower
expansion velocity
but temperatures higher than of other HVG SNe (\citealt{benetti05,tanaka08}). 
 These few events can be considered as transitional objects, linking
all three sub-classes and providing continuity between the
groups. We may also include SN 1994D in these objects as the 
top panel in Fig.~\ref{rcsix1} indicates; in addition, SN 2005hk seems to
establish a link between the LVG and FAINT SNe. Similar plots like in 
Fig.~\ref{rcsix1} of $EW_w\lbrace\ion{Si}{II}\rbrace$ versus $\Delta
m_{15}$ or stretch confirm a strong affinity of this supernova for the
FAINT group.\\
Another object of interest is the peculiar SN 2006gz (an SN with the
slowest fading light-curves ever seen in a SN~Ia), whose properties
deviate from the LVG SNe. Its early-time \ion{Si}{II} velocity is
low, like for LVG SNe, which is attributed to an envelope of
unburned carbon that slows expansion (see
~\citealt{hicken07}).

A linear fit in upper plot in Fig.~\ref{rcsix1}, with only LVG
SNe at low-$z$ included, gives:

\begin{equation*}
 EW_w\lbrace\ion{Si}{II}\rbrace= (12.21\pm0.32) -
(5.88\pm0.30)x_1.
\end{equation*} 

\begin{figure}[!htb]
\centerline{\
\includegraphics[width=8cm,angle=-90]{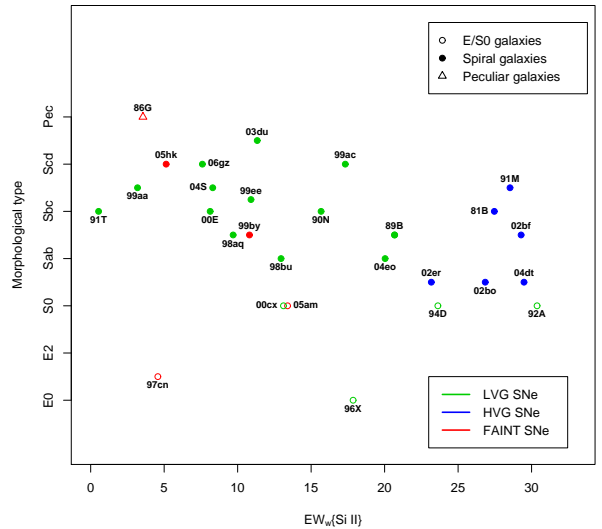}}
\caption[]{\label{rcsigalmorph}{\small\sl Morphological type of the SN host
galaxy versus measured $EW_w\lbrace\ion{Si}{II}\rbrace$ for nearby SNe.}
\bigskip}
\end{figure}

We looked for the SN host galaxy type to explore its effect on the SN
sub-class. It was emphasized by
~\citet{hamuy96,altavilla04,gallagher05} that intrinsically faint events
rather occur in E/S0 galaxies.
The diagram in Fig.~\ref{rcsigalmorph} 
points to the conclusion that early-type (E/S0) galaxies
lack SNe with smaller
$EW_w\lbrace\ion{Si}{II}\rbrace$ values, say $\leq 12$. 
Among our sample, no HVG SN
hosted in the elliptical galaxy is found, though 
there are two ambiguous SNe between LVG and HVG group in S0 galaxies. 
These two are identified as SNe 1992A and 1994D. 

Regarding the issue of whether or not the distance
estimator varies across
different environments, \citet{conley06} suggest that
stretch correction works well no matter the SN environment.
On the other hand, it is not the case with colour, since this correction
accounts for dust or extinction effects and also for the intrinsic SN
relationship (see, e.g. \citealt{sullivan06,conley07}).

\subsection{High-\texorpdfstring{$z$}{} SNe}

The method presented in this work allows us to examine in a consistent way 
low and high-$z$ SNe for which we calculated
$EW_w\lbrace\ion{Si}{II}\rbrace$ and SALT2 light-curve parameters.

Adding high-$z$ SNe to the cluster analysis 
one obtains the following SNe that are classified as HVG SNe: 03D4fd,
04D3dd, 04D3gx, b013 and d058.

Two outliers with the largest $x_1$ values can also be noticed in upper panel of Fig.~\ref{rcsix1}, namely
SCP SN 1997S 
and ESSENCE d033. In the opposite region of lower $x_1$ values,
besides SN e138, there are a few candidates, SCP 1997ai 
SNLS-03D4cz, 03D4cn and ESSENCE f216, e029 that
seem to have affinities towards the group of FAINT SNe. 

ESSENCE SN d117 is identified in the
ambiguous zone between FAINT and LVG SNe. Further, it is hard to 
classify SNe 03D1fq, 03D4fd, e132 and f011.

We do not exclude the possibility that
dispersion seen in upper plot in Fig.~\ref{rcsix1} is due to host galaxy
contamination, especially for the SNLS and ESSENCE SNe.
As it is  mentioned in \ref{systerrors} the presence of galaxy light in the 
SN spectra might lead to underestimated values of $EW_w\lbrace\ion{Si}{II}\rbrace$.

The bottom panel in Fig.~\ref{rcsix1} indicates that the correlation
between $EW_w\lbrace\ion{Si}{II}\rbrace$ and $c$ is quite weak, but there
are outliers. It can justify an eventual colour cut, say
$-0.2<c<0.2$, that would ensure a more homogeneous SN data set for
cosmological use. Indeed, SNe with $c$ out of the mentioned interval tend to 
have more dispersion in the Hubble diagram (see
~\citealt{arsenijevic08,kowalski08}).

In addition, we find no significant correlation between
$EW_w\lbrace\ion{Si}{II}\rbrace$ and the absolute magnitude $M_B$ corrected
for $x_1$ and $c$, using updated values for $\alpha$ and $\beta$ from
~\citealt{guy07}, although there are many outliers present. The linear
factor in the relation $M_{B_{corr}} \sim EW_w\lbrace\ion{Si}{II}\rbrace$
for the low-$z$ SNe is of the order of 0.01. We used values of the distance
modulus $\mu$ for low-$z$ SNe from ~\citet{tammann08}. These
questions are further discussed in ~\citet{arsenijevic08}.

A source of systematic errors in the Hubble diagram comes from
uncertainties in distance or host galaxy reddening or from
intrinsic properties of the SNe. With the available data we are not yet
able to distinguish
between potential systematic effects that come from magnitude
dependence on intrinsic colour and the colour contribution due to dust
effects in the host galaxy of the SNe. Note that in order to examine the
dust properties of distant galaxies using SNe~Ia, one needs to first 
remove the intrinsic dependence of SNe luminosity on the light-curve
parameters.\\

An important issue in cosmology with SNe Ia is how to clean the SN sample 
of any objects that might not obey a standard peak 
luminosity/light-curve shape relation. 
Our results confirm the possible contamination of the high-$z$ sample with 
HVG SNe. This fact deserves further attention as a potential source of
systematic error, 
as was also suggested in ~\citealt{tanaka08}.

Previous comparisons between low and high-$z$ SNe spectral features
have shown the homogeneity within the two samples.
We have taken a step further by introducing a new approach that allows us 
to identify SN Ia sub-classes, independent of the SN redshift, 
using only one spectrum with \ion{Si}{II} 4000  together with
the light-curve width and colour parameters.

\section{Conclusions}

The use of SNe~Ia for cosmology relies on empirical calibration
techniques on the light-curves and K-correction. It
would be reassuring to
have a usable indicator measured directly from SN spectra that can be
used as an independent calibrator of the luminosity of an SN Ia event.
Applying a straightforward transformation using wavelets, we were able to
estimate, in a consistent manner on a fair number of SNe~Ia, the
equivalent width of the \ion{Si}{II} 4000 feature which previously showed
potential use for cosmology. At low redshift, we were able to
automatically distinguish three classes of SNe~Ia previously found by
other authors using much more information, such as the expansion velocity gradient.

The same wavelet-based  approach was applied to high redshift data, revealing an 
analogous clustering tendency
to that found for low redshift supernovae. 
Yet it is not clear whether the three sub-classes, found in the low-$z$ and
verified  in the high-$z$ sample,  
are completely distinct or come from one continuous family. If so,  this  could possibly be the
remnant of an extra  parameter or a more complex modelling of the
supernova data.
The implementation of our  method using recently available much larger 
spectral samples and  the use of different SN Ia sub-classes in cosmological analysis
is now in preparation.

Although the method presented here is promising and we are planning to apply it to
other spectral features,  we remark that line ratios,
pseudo-equivalent widths and their current derivatives are
not optimal to
extract information on lower signal-to-noise spectra. Parameters calculated from 
spectral features, such as equivalent widths, still consider only local 
information and each feature independently. 
New indicators based on combining local information, such as wavelet coefficients 
from the whole spectra, can provide a better characterisation of the
supernovae.

\begin{acknowledgements}
This work was supported by Funda\c{c}\~ao para a Ci\^encia e
Tecnologia (FCT), Portugal under POCTI/CTE-AST/57664/2004.  V.
Arsenijevic acknowledges support from FCT under grant no.
SFRH/BD/11119/2002 and S. Fabbro grant no. SFRH/BPD/31817/2006. Most of the
code was written in 
{\tt R}\footnote{R
Development Core Team (2007). R: A language and environment for statistical
computing. R Foundation for Statistical Computing, Vienna, Austria. ISBN
3-900051-07-0, URL \protect\url{http://www.r-project.org}}.
We would
like to thank G. Nason for useful comments on the implementation of
{\it wavethresh} package.
\end{acknowledgements}

\onecolumn
\begin{center}
\setlongtables
\begin{longtable}[c]{|l|c|c|c|}
\hline \hline
\textbf{\small SN name}&
\textbf{\small Epoch}&
\textbf{\small Bands}&
\textbf{\small References}\\
\textbf{{ }}&
\textbf{{\small (with respect to $B$ maximum)}}&
\textbf{{ }}&
\textbf{{ }}\\[0.5ex]
\endfirsthead



\hline\hline
1981B   & 0 & UBV  & 1\\
1986G   &  -3, -1, 0, 1, 2 & BV  & 2\\
1989B   & -5, -1, 0, 3, 5 & UBVRI  & 3, 4\\
1990N   & 2, 7 & UBVRI  & 5\\
1991M   & 3 & VRI  & 6\\
1992A   & -5, -1, 3, 5 & UBVRI & 7\\
1994D   & -5, -4, -2, 2, 4 & UBVRI  & 8\\
1996X   & -2,  0, 1 & BVRI  & 9\\
1998aq  &  0, 1, 2, 3, 4, 5 &  UBVRI  & 10\\
1998bu  & -3, -2, -1 & UBVRI  & 11\\
1999ee  & -4, -2, 0, 5 & UBVRI  & 12\\
2000E   & -3, -2 & UBVRI  & 13\\
2002bf  &  3 & BVRI  & 14\\
2002bo  & -5, -4, -3, -2, -1, 4 & UBVRI  & 15\\
2002er  & -5, -4, -3, -2, -1, 0, 2, 4, 5 & UBVRI & 16, 17 \\
2003cg  & -5, -1, 1, 4 & UBVRI  & 18\\
2003du  & -4, -2, -1, 0, 1, 3, 4 & UBVR  & 19, 20\\
2004dt  & -5, -4, -3, -2, -1, 0, 1, 2, 3, 4 & $\cdots$ & 14, 21\\
2004eo  & -3, 2 & UBVRI  & 22\\
2004S   &  1 & UBVRI & 23\\
2005am  & -1, 2 & UBV  & 24\\
2005cf  & -4, -3, -1, 0, 4, 5 & UBVRI & 25, 26\\
2005hj  & -5, 0, 2, 3, 5 & $\cdots$ & 27\\
2006gz  & -4, -1 & UBVri  & 28\\
\hline
\noalign{\smallskip}
\multicolumn{4}{c}{\textbf{\small SNe 91bg-like}}\\
\noalign{\smallskip}
\hline
1991bg   & 1, 2 & BVRI  & 29\\
1999by   & -5, -4, -3, 3, 4, 5 & UBVRI  & 30\\
\hline
\noalign{\smallskip}
\multicolumn{4}{c}{\textbf{\small SNe 91T-like}}\\
\noalign{\smallskip}
\hline
1991T   & -3, 0 & UBVRI  &  31, 32, 33\\
2000cx  & -3, -2, 0, 1, 5 & UBVRI  & 34\\ %
\hline
\noalign{\smallskip}
\multicolumn{4}{c}{\textbf{\small SNe Ia peculiar}}\\
\hline
1997br  & -4 & UBVRI  & 35\\
1997cn  &  3 & UBVRI  & 36\\ 
1999aa  & -3,-1 & UBVRI  & 37\\
1999ac  &  0, 2 & UBVRI  & 38, 39\\
1999aw  &  3, 5 & BVRI  & 40\\
2002cx  & -4, -1 & BVRI  & 41, 42\\
2005hk  & -5, -4, 4 & UBVRI  & 43, 44\\
\hline
\noalign{\smallskip}
\multicolumn{4}{c}{ }\\
\caption[The list of nearby SNe]{\small\sl List of 
nearby SNe used in this work.\\ \\
{\footnotesize References: (1) \citealt{branch83}; 
(2) \citealt{phillips87}; (3) \citealt{barbon90}; (4) \citealt{wells94};
(5) \citealt{mazzali93}; (6) \citealt{gomez98}; (7) \citealt{kirshner93};
(8) \citealt{patat96}; (9) \citealt{salvo01}; (10) \citealt{branch03}; 
(11) \citealt{jha99}; (12) \citealt{hamuy02}; (13) \citealt{valentini03}; 
(14) \citealt{leonard05}; (15) \citealt{benetti04};  
(16) \citealt{pignata04}; (17) \citealt{kotak05};
(18) \citealt{eliasrosa06}; (19) \citealt{anupama05}; 
(20) \citealt{stanishev07}; (21) \citealt{altavilla07}; 
(22) \citealt{pastorello07b}; (23) \citealt{krisciunas07}; 
(24) \citealt{brown05}; 
(25) \citealt{garavini07b}; (26) \citealt{pastorello07a}; 
(27) \citealt{quimby07}; (28) \citealt{hicken07}; 
(29) \citealt{turatto96}; (30) \citealt{garnavich04}; 
(31) \citealt{jeffery92}; (32) \citealt{phillips92}; 
(33) \citealt{mazzali95}; (34) \citealt{li01}; (35) \citealt{li99}; 
(36) \citealt{turatto98}; (37) \citealt{garavini04}; 
(38) \citealt{garavini05}; (39) \citealt{phillips06}; 
(40) \citealt{strolger02}; (41) \citealt{li03}; (42) \citealt{branch04};
(43) \citealt{phillips07}; (44) \citealt{sahu08}}}\label{nearbylistaa2}

\end{longtable}
\end{center}
\vspace{0.5cm}
\begin{center}
\setlongtables
\begin{longtable}[c]{|l|c|r|r|r|r|}
\hline \hline
\textbf{\small SN name}&
\textbf{\small $z_{\textrm{CMB}}$}&
\textbf{\small $m_B$}&
\textbf{\small $x_1$}&
\textbf{\small $c$}&
\textbf{\small $EW_w\lbrace\ion{Si}{II}\rbrace$}\\[0.5ex]
\endfirsthead
\hline\hline

\hline
\textbf{{\small SN name}}&
\textbf{{\small $z_{\textrm{CMB}}$}}&
\textbf{\small $m_B$}&
\textbf{\small $x_1$}&
\textbf{\small $c$}&
\textbf{{\small $EW_w\lbrace\ion{Si}{II}\rbrace$}}\\[0.5ex]\hline
\endhead
\hline
\multicolumn{6}{c}{ }\\
\multicolumn{6}{l}{{Continued on next page\ldots}}\\
\endfoot
\endlastfoot

\hline\hline
1981B  & $0.0072$& $12.0074\pm0.0217$ & $-0.6660\pm0.1673$ & $
0.1530\pm0.0196$ & $27.47\pm2.33$\\
1986G  & $0.0028$& $12.0184\pm0.0371$ & $-3.0859\pm0.3261$ & $
0.9851\pm0.0409$ & $3.56\pm2.67$\\
1989B  & $0.0036$& $12.2424\pm0.0177$ & $-1.0569\pm0.1418$ & $
0.4736\pm0.0141$ & $20.67\pm1.48$\\
1990N  & $0.0045$& $12.6599\pm0.0138$ & $ 0.7208\pm0.1168$ & $
0.0773\pm0.0121$ & $15.68\pm2.45$\\
1991M  & $0.0076$& $14.3553\pm0.1376$ & $-1.5904\pm0.3325$ & $
0.0036\pm0.1140$ & $28.53\pm2.31$\\
1992A  & $0.0059$& $12.5356\pm0.0100$ & $-2.0636\pm0.0995$ & $
0.0810\pm0.0103$ & $30.37\pm2.42$\\
1994D  & $0.0027$& $11.7314\pm0.0095$ & $-2.1674\pm0.0842$ & $
-0.0643\pm0.0100$& $23.62\pm2.47$\\
1996X  & $0.0078$& $12.9832\pm0.0151$ & $-1.2519\pm0.1539$ & $
0.0408\pm0.0153$ & $17.86\pm2.61$\\
1998aq & $0.0042$& $12.2893\pm0.0070$ & $-0.4507\pm0.0615$ &
$-0.0920\pm0.0076$ & $9.69\pm2.52$\\
1998bu & $0.0042$& $12.0777\pm0.0062$ & $-0.3570\pm0.0556$ & $
0.3320\pm0.0068$ & $12.95\pm1.37$\\
1999ee & $0.0105$& $14.8268\pm0.0062$ & $ 0.4171\pm0.0526$ & $
0.3052\pm0.0067$ & $10.91\pm2.09$\\
2000E  & $0.0048$& $12.8034\pm0.0099$ & $ 0.4851\pm0.1117$ & $
0.2058\pm0.0101$ & $8.13\pm2.47$\\
2002bf & $0.0247$& $16.2978\pm0.0445$ & $-0.4205\pm0.2040$ & $
0.2293\pm0.0275$ & $29.29\pm2.32$\\
2002bo & $0.0053$& $13.9533\pm0.0086$ & $-0.6595\pm0.0668$ & $
0.4711\pm0.0090$ & $26.85\pm2.35$\\
2002er & $0.0086$& $14.2263\pm0.0100$ & $-1.1062\pm0.0825$ & $
0.1918\pm0.0102$ & $23.18\pm2.32$\\
2003cg & $0.0041$& $15.7912\pm0.0089$ & $-0.4629\pm0.0655$ & $
1.3005\pm0.0091$ & $20.69\pm2.40$\\
2003du & $0.0064$& $13.4923\pm0.0093$ & $ 0.1035\pm0.0697$ &
$-0.0472\pm0.0087$ & $11.33\pm0.94$\\
2004dt & $0.0188$& $\cdots$ & $\cdots$ & $\cdots$ & $29.48\pm2.35$\\
2004eo & $0.0147$& $15.0554\pm0.0066$ & $-1.4892\pm0.0601$ & $
0.1191\pm0.0075$ & $20.03\pm4.19$\\
2004S  & $0.0091$& $14.1485\pm0.0293$ & $-0.3168\pm0.1622$ & $
0.1317\pm0.0166$ & $8.30\pm2.58$\\
2005am & $0.0090$& $13.7303\pm0.0225$ & $-2.6235\pm0.1490$ & $
0.2102\pm0.0167$ & $13.40\pm3.09$\\
2005cf & $0.0070$& $13.0841\pm0.0054$ & $-0.3598\pm0.0471$ & $
-0.0387\pm0.0060$ & $14.70\pm1.53$\\
2005hj & $0.0574$& $\cdots$ & $\cdots$ & $\cdots$ & $8.36\pm2.76$\\
2006gz & $0.0277$& $15.7957\pm0.0176$ & $ 2.3344\pm0.1741$ & $
0.0490\pm0.0148$ & $7.60\pm4.17$ \\
\hline\hline
\noalign{\smallskip}
\multicolumn{6}{c}{\textbf{\small SNe 91bg-like}}\\
\noalign{\smallskip}
\hline
1991bg  & $0.0042$& $14.6012\pm0.0381$ & $-2.7372\pm0.1940$ & $
0.7185\pm0.0309$ & $\cdots$\\
1999by  & $0.0029$& $13.6486\pm0.0143$ & $-3.0586\pm0.2231$ & $
0.5483\pm0.0148$ & $10.80\pm4.38$\\
\hline\hline
\noalign{\smallskip}
\multicolumn{6}{c}{\textbf{\small SNe 91T-like}}\\
\noalign{\smallskip}
\hline
1991T  & $0.0070$& $11.5450\pm0.0118$ & $ 1.0398\pm0.1138$ & $
0.2008\pm0.0121$ & $0.54\pm1.77$\\
2000cx & $0.0079$& $13.0444\pm0.0082$ & $-1.3500\pm0.0790$ & $
0.0604\pm0.0090$ & $13.12\pm2.92$\\ %
\hline\hline
\noalign{\smallskip}
\multicolumn{6}{c}{\textbf{\small SNe Ia peculiar}}\\
\hline
1997br & $0.0080$& $13.4941\pm0.0160$ & $ 0.0209\pm0.1319$ & $
0.2849\pm0.0128$ & $\cdots$\\
1997cn & $0.0170$& $15.9641\pm0.1168$ & $-2.6791\pm0.3456$ & $
0.1092\pm0.0753$ & $4.57\pm3.24$\\
1999aa & $0.0144$& $14.7077\pm0.0131$ & $ 1.1371\pm0.1313$ &
$-0.0242\pm0.0124$ & $3.19\pm2.66$\\
1999ac & $0.0099$& $14.1078\pm0.0092$ & $-0.1542\pm0.1064$ & $
0.1118\pm0.0102$ & $17.33\pm2.35$\\
1999aw & $0.0393$& $16.7145\pm0.0163$ & $ 2.0927\pm0.1799$ & $-0.0069\pm
0.0157$ & $1.08\pm1.54$\\
2002cx & $0.0250$& $17.7281\pm0.0189$ & $ 0.1253\pm0.1507$ & $
0.3253\pm0.0154$ & $\cdots$\\
2005hk & $0.0118$& $15.9015\pm0.0112$ & $-0.4887\pm0.1149$ & $
0.3112\pm0.0111$ & $5.13\pm2.53$\\
\hline
\noalign{\smallskip}
\multicolumn{6}{c}{ }\\
\caption{\small\sl SALT2 estimates and $EW_w\lbrace\ion{Si}{II}\rbrace$
results for nearby SNe. The blank spaces ``$\cdots$'' in
$EW_w\lbrace\ion{Si}{II}\rbrace$ column indicate insufficient presence of
\ion{Si}{II} feature despite an adequate wavelength coverage.}
\label{nearbysalt2}
\end{longtable}
\end{center}

\vspace{1.5cm}
\begin{center}
\setlongtables
\begin{longtable}[c]{|l|l|r|r|r|r|c|}
\hline \hline
\textbf{\small SN name}&
\textbf{\small $z_{\textrm{CMB}}$}&
\textbf{\small $m_B$}&
\textbf{\small $x_1$}&
\textbf{\small $c$}&
\textbf{\small $EW_w\lbrace\ion{Si}{II}\rbrace$}&
\textbf{\small Reference}\\[0.5ex]
\endfirsthead
\hline\hline

\hline
\textbf{\small SN name}&
\textbf{\small $z_{\textrm{CMB}}$}&
\textbf{\small $m_B$}&
\textbf{\small $x_1$}&
\textbf{\small $c$}&
\textbf{\small $EW_w\lbrace\ion{Si}{II}\rbrace$}&
\textbf{\small Reference}\\[0.5ex]\hline
\endhead
\hline
\multicolumn{7}{c}{ }\\
\multicolumn{7}{l}{{Continued on next page\ldots}}\\
\endfoot
\endlastfoot

\hline\hline
03D1ax  &  0.496  & $22.9691\pm0.0160$ & $-1.0045\pm0.1715$ & $-0.0304\pm0.0285$ &  $9.01\pm3.95$ & 1\\
03D1bk  &  0.865  &  $\cdots$   &  $\cdots$ & $\cdots$ & $13.50\pm5.36$ &
1\\
03D1co  &  0.68   & $24.1069\pm0.0481$ & $ 0.9602\pm0.6335$ &
$-0.0229\pm0.0539$ &  $8.21\pm5.94$ & 1\\  
03D1ew  &  0.868  & $24.3489\pm0.0576$ & $ 0.4624\pm0.4097$ &
$-0.0762\pm0.1830$ & $ 9.94\pm8.66$ & 1\\
03D1fq  &  0.80   & $24.5299\pm0.0343$ & $-1.8585\pm0.5215$ &
$-0.2695\pm0.1905$ & $31.30\pm4.38$ & 1\\
03D4cn  &  0.818  & $24.6580\pm0.0401$ & $-2.3380\pm0.6214$ & $
0.1416\pm0.1758$ & $17.25\pm4.15$ & 1\\
03D4cy  &  0.9271 & $24.6993\pm0.0873$ & $ 0.2584\pm0.4974$ &
$-0.3657\pm0.1862$ & $21.11\pm8.40$ & 1\\
03D4cz  &  0.695  & $24.0406\pm0.0413$ & $-2.3515\pm0.4144$ &
$-0.0699\pm0.0605$ & $ 6.09\pm4.62$ & 1\\
03D4fd  &  0.791  & $24.2177\pm0.0248$ & $-0.1670\pm0.4221$ &
$-0.0418\pm0.0527$ & $23.54\pm6.08$ & 1\\
03D4gl  &  0.56   & $23.1869\pm0.0428$ & $-0.8379\pm0.5806$ &
$-0.0111\pm0.0426$ & $ 5.58\pm6.56$ & 1\\
04D1de  &  0.7677 & $\cdots$   &  $\cdots$ & $\cdots$ & $ 5.73\pm5.24$ &
1\\
04D1hd  &  0.3685 & $\cdots$   &  $\cdots$ & $\cdots$ & $15.86\pm3.87$ &
1\\
04D3dd  &  1.01   & $25.0592\pm0.1625$ & $ 1.1167\pm0.6553$ &
$-0.1155\pm0.2196$ & $44.42\pm8.38$ & 1\\
04D3fq  &  0.73   & $24.1009\pm0.0262$ & $-0.7710\pm0.4011$ &
$-0.0168\pm0.0480$ & $14.31\pm7.25$ & 1\\
04D3gx  &  0.91   & $24.6957\pm0.0732$ & $-0.1806\pm0.5141$ &
$-0.1816\pm0.1746$ & $29.07\pm10.80$ & 1\\
04D3kr  &  0.3373 & $21.9321\pm0.0207$ & $ 1.0967\pm0.1851$ & $
0.0131\pm0.0167$ & $10.69\pm4.27$ & 1\\
04D3lp  &  0.983  & $24.8964\pm0.1414$ & $-1.4766\pm0.5405$ & $
0.1080\pm0.2277$ & $20.86\pm6.11$ & 1\\
04D3mk  &  0.813  & $\cdots$   &  $\cdots$ & $\cdots$ & $ 4.70\pm4.37$ &
1\\
04D3ml  &  0.95   & $24.5060\pm0.0725$ & $ 1.4300\pm0.7394$ & $
0.0824\pm0.1447$ & $10.99\pm8.40$ & 1\\
04D3nh  &  0.3402 & $22.0925\pm0.0161$ & $ 0.5018\pm0.1758$ & $
0.0350\pm0.0148$ & $20.07\pm4.24$ & 1\\
04D3ny  &  0.81   & $24.3118\pm0.0418$ & $ 2.4869\pm1.2655$ &
$-0.1175\pm0.1758$ & $ 8.91\pm6.51$ & 1\\
04D4dm  &  0.811  & $24.3974\pm0.0404$ & $ 0.8626\pm1.1636$ &
$-0.1282\pm0.1661$ & $11.50\pm7.09$ & 1\\
04D4hu  &  0.7027 & $\cdots$   &  $\cdots$ & $\cdots$ &  $14.21\pm3.89$ &
1\\
04D4ic  &  0.68   & $\cdots$   &  $\cdots$ & $\cdots$ &  $15.33\pm4.63$ &
1\\
04D4ii  &  0.866  & $\cdots$   &  $\cdots$ & $\cdots$ &  $14.24\pm4.74$ &
1\\
04D4jy  &  0.93   & $\cdots$   &  $\cdots$ & $\cdots$ &  $22.66\pm11.59$ &
1\\
b010    &  0.591 & $23.4054\pm0.0790$ & $ 1.9986\pm0.9094$ &
$-0.1171\pm0.0948$ & $ 6.16\pm3.17$ & 2\\
b013    &  0.426 & $22.6230\pm0.0427$ & $-0.5961\pm0.4047$ & $
0.0244\pm0.0524$ & $32.85\pm2.88$ & 2\\
b016    &  0.329 & $22.5340\pm0.1385$ & $ 0.2256\pm0.7683$ & $
0.2086\pm0.1276$ & $ 7.50\pm3.21$ & 2\\
b020    &  0.425  & $22.4605\pm 0.2181$ & $-1.4169\pm 1.8301$ &
$0.0434\pm0.1987$ & $13.51\pm2.72$ & 2\\
d033    &  0.531 & $23.0873\pm0.0802$  & $3.8152\pm 1.1886$ &
$-0.2914\pm 0.0986$ & $18.13\pm3.92$ & 2\\
d058    &  0.583 & $23.5307\pm0.0418$ & $ 0.1674\pm0.6575$ & $
0.1711\pm0.0754$ & $36.00\pm3.91$ & 2\\
d083    &  0.333 & $21.0031\pm0.0333$ & $  1.2251\pm 0.2224$ &
$-0.0512\pm 0.0349$ & $9.39\pm2.64$ & 2\\
d085    &  0.401 & $22.4501\pm0.0480$ & $ 0.6541\pm0.3903$ & $
0.0428\pm0.0590$ & $13.87\pm2.99$ & 2\\
d093    &  0.363 & $21.8967\pm0.0480$ & $ 1.0120\pm0.2783$ &
$-0.0994\pm0.0482$ & $ 7.89\pm7.35$ & 2\\
d117    &  0.309 & $22.3667\pm0.0997$ & $-1.8445\pm0.3583$ & $
0.0968\pm0.0916$ & $21.98\pm2.76$ & 2\\
d149    &  0.342 & $22.0957\pm0.0455$ & $ 0.3705\pm0.3584$ & $
0.0109\pm0.0436$ & $ 9.52\pm3.10$ & 2\\
e020    &  0.159 & $21.0616\pm0.1498$ & $  0.0291\pm 0.3768$ & $ 0.3722\pm
0.1266$ & $13.18\pm2.35$ & 2\\
e029    &  0.332 & $22.4054\pm0.0803$ & $-2.3762\pm1.0506$ & $
0.0532\pm0.0810$ & $15.36\pm2.60$ & 2\\
e108    &  0.469 & $22.5728\pm0.0815$ & $ 2.0057\pm1.5990$ &
$-0.1079\pm0.0887$ & $14.85\pm3.32$ & 2\\
e132    &  0.239 & $21.7710\pm0.0645$ & $-0.3371\pm0.2480$ & $
0.3003\pm0.0605$ & $21.33\pm2.47$ & 2\\
e138    &  0.612 & $23.7475\pm0.0501$ & $-3.8738\pm0.7231$ & $
0.2588\pm0.0996$ & $18.40\pm3.21$ & 2\\
e147    &  0.645 & $23.3283\pm0.0463$ & $-0.3841\pm0.4340$ &
$-0.1072\pm0.0570$ & $18.86\pm2.90$ & 2\\
e148    &  0.429 & $22.6329\pm0.0335$ & $-1.1050\pm0.3499$ &
$-0.0489\pm0.0448$ & $15.42\pm2.66$ & 2\\
f011    &  0.539 & $23.2794\pm0.0363$ & $-1.4498\pm0.4991$ & $
0.0598\pm0.0620$ & $23.81\pm2.91$ & 2\\
f041    &  0.561 & $23.1138\pm0.0479$ & $ 2.4526\pm1.2001$ &
$-0.0763\pm0.0650$ & $15.07\pm3.00$ & 2\\
f076    &  0.410 & $22.3460\pm0.0641$ & $-1.0001\pm0.6384$ & $
0.0600\pm0.0899$ & $15.94\pm3.44$ & 2\\
f096    &  0.412 & $22.8398\pm0.2053$ & $ 0.1328\pm2.0466$ & $
0.2771\pm0.1320$ & $ 8.35\pm2.76$ & 2\\
f216    &  0.599 & $23.7736\pm0.0775$ & $-3.2479\pm1.0666$ &
$-0.0383\pm0.1316$ & $12.44\pm3.15$ & 2\\
f231    &  0.619 & $23.4465\pm0.0355$ & $ 1.2394\pm0.3607$ &
$-0.0759\pm0.0623$ & $17.03\pm3.30$ & 2\\
f235    &  0.422 & $22.4017\pm0.0552$ & $-1.1912\pm0.4730$ &
$-0.0813\pm0.0652$ & $14.07\pm2.46$ & 2\\
f244    &  0.540 & $23.2981\pm0.0524$ & $-0.5891\pm0.7566$ &
$-0.0055\pm0.0801$ & $16.24\pm3.27$ & 2\\
1997G   &  0.763  & $24.2807\pm0.2579$ & $-0.8583\pm1.3793$ &
$-0.4161\pm0.2310$ & $ 7.08\pm3.37$ & 3\\
1997S   &  0.612  & $23.4756\pm0.0639$ & $ 3.6292\pm1.7033$ & $
0.1103\pm0.1169$ & $21.70\pm2.72$ & 3\\
1997af  &  0.579  & $23.5343\pm0.0796$ & $-1.0871\pm0.4901$ &
$-0.2081\pm0.1852$ & $ 7.78\pm3.04$ & 3\\
1997ai  &  0.454  & $22.8788\pm0.0627$ & $-2.6928\pm0.9158$ &
$-0.0040\pm0.0787$ & $21.42\pm2.74$ & 3\\
1997aj  &  0.581  & $23.1959\pm0.0822$ & $ 0.7138\pm0.7974$ & $
0.1991\pm0.1123$ & $ 6.23\pm2.60$ & 3\\
1997ap  &  0.831  & $24.3496\pm0.0633$ & $ 0.3579\pm0.5601$ &
$-0.0115\pm0.0384$ & $17.21\pm3.82$ & 3\\
2000fr  &  0.543  & $23.0376\pm0.0263$ & $ 0.7208\pm0.2533$ &
$-0.0359\pm0.0281$ & $ 7.81\pm2.70$ & 4\\
2001gm  & 0.478   & $\cdots$   &  $\cdots$ & $\cdots$ &  $15.52\pm7.74$ &
4\\ 
2001go  & 0.552   & $\cdots$   &  $\cdots$ & $\cdots$ &  $27.15\pm4.34$ &
4\\
2001gw  & 0.363   & $\cdots$   &  $\cdots$ & $\cdots$ &  $ 9.57\pm3.77$ &
4\\
2001gy  & 0.511   & $\cdots$   &  $\cdots$ & $\cdots$ &  $ 3.66\pm3.23$ &
4\\
2001ha  & 0.58    & $\cdots$   &  $\cdots$ & $\cdots$ &  $21.80\pm5.40$ &
4\\
2001hc  & 0.35    & $\cdots$   &  $\cdots$ & $\cdots$ &  $ 7.28\pm2.93$ &
4\\
2002gl  & 0.510   & $\cdots$   &  $\cdots$ & $\cdots$ &  $16.25\pm3.30$ &
4\\
2002km  & 0.606   & $\cdots$   &  $\cdots$ & $\cdots$ &  $15.05\pm4.37$ &
4\\
2002ks  & 1.181   & $\cdots$   &  $\cdots$ & $\cdots$ &  $20.79\pm6.45$ &
4\\
1999ff  &  0.455  & $23.2026\pm0.0516$ & $-1.4400\pm0.6801$ &
$-0.0109\pm0.0698$ & $21.45\pm8.25$ & 5\\
1999fh  & 0.369   & $\cdots$   &  $\cdots$ & $\cdots$ & $18.47\pm2.76$ &
5\\
1999fj  &  0.816  & $24.1845\pm0.0621$ & $ 0.2386\pm0.5735$ & $-0.0833\pm0.1807$ & $16.82\pm3.33$ & 5\\
1999fk  &  1.057  & $24.7553\pm0.0609$ & $-0.2806\pm0.9992$ & $-0.0586\pm0.0707$ & $14.60\pm6.08$ & 5\\
1999fm  &  0.950  & $24.2363\pm0.0445$ & $ 1.5816\pm0.4610$ & $ 0.0279\pm0.0904$ & $ 5.81\pm7.05$ & 5\\
1999fn  &  0.477  & $22.7513\pm0.0393$ & $ 0.7821\pm0.3983$ & $ 0.0230\pm0.0415$ & $12.51\pm2.98$ & 5\\
1999fw  &  0.278  & $21.6893\pm0.0671$ & $-1.2738\pm0.6348$ & $ 0.0975\pm0.0616$ & $12.85\pm2.84$ & 5\\
\hline
\noalign{\smallskip}
\multicolumn{7}{c}{ }\\
\caption[]{\small\sl SALT2 estimates and
  $EW_w\lbrace\ion{Si}{II}\rbrace$ for high-z sample.\\ \\
{\footnotesize References: (1) \citealt{howell05}; (2)
\citealt{matheson05}; (3) \citealt{hook05}; (4) \citealt{lidman05}; (5)
\citealt{tonry03}}}
\label{salt2list}
\end{longtable}
\end{center}

\end{document}